\documentclass{appolb}
\usepackage{epsfig}

\newcommand{\cF}{{\cal F}}
\newcommand{\cE}{{\cal E}}
\newcommand{\cC}{{\cal C}}
\newcommand{\cR}{{\cal R}}
\newcommand{\barcC}{{{\bar{\cal C}}}}

\newcommand{\as}{{\alpha_s}}
\newcommand{\ee}{e^+e^-}

\newcommand{\jphg}[3]{{\em J. Phys. {\bf G #1} (#2) #3}}

\newcommand{\prd}[3]{{\em Phys. Rev. {\bf D #1} (#2) #3}}
\newcommand{\npb}[3]{{\em Nucl. Phys. {\bf B #1} (#2) #3}}
\newcommand{\jhep}[3]{{\em J. High Energy Phys.} {\bf #1} (#2) #3}
\newcommand{\plb}[3]{{\em Phys. Lett.} {\bf B #1} (#2) #3}

\newcommand{\caesar}{\textsc{caesar}\,\,}

\begin{document}
\title{Resummed event shapes at hadron colliders%
\thanks{Presented at the XXXIV International Symposium on Multiparticle Dynamics}%
} \author{Giulia Zanderighi \address{Fermi National Laboratory, P.O.
    Box 500, 60510 Batavia, IL} } \maketitle
\begin{abstract}
  We present recently defined jet-observables for hadron-hadron dijet
  production, which are designed to reconcile the seemingly
  conflicting theoretical requirement of globalness, which makes it
  possible to resum them (automatically) at NLL accuracy and the
  limited experimental reach of detectors, so that they are measurable
  at the Tevatron and at the LHC.
\end{abstract}
  
\section{Introduction}
Event shapes and jet-rates are infrared and collinear (IRC) safe
observables, which describe the energy and momentum flow of the final
state. They constitute an ideal compromise between simplicity and
sensitivity to properties of QCD radiation.  They provide then a
wealth of information, e.g.  in measurements of the coupling
$\alpha_s$ and its renormalisation group running, in cross
checks/measurements of the values of the colour factors of QCD and,
most importantly, in studies of the connection between parton-level
(the perturbative (PT) description of quarks and gluons) and
hadron-level (the real), for a review see~\cite{DasSalReview}.

IRC-safety ensures that event-shape distributions can be computed within
perturbation theory, however in the more exclusive phase space region
where perturbative radiation is suppressed (conventionally associated
to almost vanishing values of the observable, $V \ll 1$) large
logarithmic corrections need to be resummed to all orders.
More specifically, given an event-shape $V$, a function of all
secondary final state momenta $\{k_1, \dots, k_n\}$ and of the set of
recoiling Born momenta $\{\tilde p\}$, the probability of
``constrained events'' i.\ e.\ $V(\{\tilde p\}, k_{1}\dots k_n)< v$
has a {\emph{divergent} PT expansion} for $v \to 0$
$$ \Sigma(v) \equiv \mathrm{Prob} (V < v) = 1 + \sum_{m \le 2n} R_{n, m} \as^n \ln^m v + \dots\>, $$
i.\ e.\ there is a soft and collinear divergence [$\leadsto \ln v$] for each
emitted gluon. 
Today's state-of-the art accuracy accounts for all
Leading (LL) 
and Next-to-Leading Logarithms (NLL) as follows: 
\begin{equation}
\label{eq:Sigma}
\Sigma(v) = \exp\{{\underbrace{L g_1(\as L)}_{LL} +
  \underbrace{g_2(\as L)}_{NLL}}+ \dots \}\>,\qquad L \equiv
\ln\frac1v\>.
\end{equation}
Furthermore resummations are matched to fixed order results at NLO.

\section{Basics of resummation}
The main ingredient in resummations is factorisation. Usually two
steps are needed to obtain a resummed prediction.
The first task is to exploit angular ordering to reduce $n$-parton
matrix elements to a QED-like factorised form, i.\ e.\ to the product
of independent emissions from the hard scattering partons only.
Schematically, e.g.  for $e^+e^- \to 2\>\mbox{jets}$ this can be
written as
  \begin{equation}
\label{eq:Mfact}
    w_{p\bar p}(k_1,\dots,k_n)
={\frac{1}{n!} \prod_{i=1}^n} w_{p\bar p}(k_i)
\sim \frac{1}{n!} \prod_{i=1}^n 
\frac{\as C_F}{\pi}\frac{dE}{E}\frac{d\theta}{\theta}\>,
  \end{equation}
  with corrections which contribute beyond NLL to $\Sigma(v)$.  The
  second step requires some analytical understanding of the
  observable's behaviour in the presence of soft-collinear emissions
  in order to factorise its definition via Mellin transforms, i.\ g.\ 
  for the thrust $T$ in $e^+e^- \to$2 jets one has ($Q = \frac{\sqrt{s}}{2}$)
    \begin{equation}
\label{eq:Vfact}
         1-T \simeq \frac1Q {\sum_{i=1}^n} \frac{E_i\theta_i^2}{2}
        \quad {\longrightarrow} \quad 
      \Theta(1-T<\tau)= \int \frac{d\nu}{2\pi i\nu} 
      e^{\nu \tau} {\prod_{i=1}^n}
      e^{-\nu \frac{E_i\theta_i^2}{2Q}}\>.
    \end{equation}
    By combining eq.~(\ref{eq:Mfact}) and eq.~(\ref{eq:Vfact}) one
    obtains the resummed answer
\begin{equation}
\label{eq:Tres}
\Sigma(\tau) \int \frac{d\nu}{2\pi i\nu} e^{\nu \tau} \exp\left[
  \int \frac{d\theta}{\theta} \frac{dE}{E} \frac{\as(E
    \theta)C_F}{\pi} \left(e^{-\nu \frac{E_i\theta_i^2}{2Q}} -
    1\right)\right]\>,
\end{equation}
where the effects due to the running of the coupling have been
included.

Despite the simplicity of the ideas underlying resummations, some
technicalities cannot be avoided.
Indeed, at NLL accuracy, care is needed to treat properly the emission
of hard collinear or soft large-angle gluons and the inclusive gluon
splittings.
Furthermore, it is known that for some observables, when seeking NLL
accuracy, the multi-parton matrix-element cannot be factorised as in
eq.~(\ref{eq:Mfact}), notably this is the case for non-global
observables~\cite{NG1}.  In a similar way some observables, such as
jet rates in Jade-like algorithms~\cite{JadeDL}, do not exponentiate,
so that it is not possible to write the distribution in the form in
eq.~(\ref{eq:Sigma}).  Also, the observable's factorisation itself can
be non trivial especially for multi-jet observables like the $3$-jet
limit of thrust minor, which involves 5 integral
transforms~\cite{eeKout}.
\section{Automated resummed predictions}
The observation that the origin of logarithmic enhancement in
jet-shape distributions is always the same (i.\ e.\ it is due to the
radiation of soft-collinear gluons) and that the approximations which
lead to resummed expressions such as eq.~(\ref{eq:Tres}) are very
similar for different observables, it was natural to investigate the
possibility of developing a general and automated approach to
resummation, which does not require analytical treatment of observable
specific multiple emission effects via Mellin/Fourier transforms.

Such an approach is feasible, and has been implemented in the computer
code \caesar (Computer Automated Expert Semi-Analytical
Resummer).~\cite{Caesar} This program differs from usual Monte Carlos
whose task is generally the numerical evaluation of integrals via
generation of random events.
\caesar instead is based on a generic master resummation formula,
presented and derived in detail in~\cite{Caesar}. This formula applies
to a well-defined class of observables, whose requirements can be
summarised as follows.
\begin{itemize}
\item Given a Born event, when just one soft emission $k$ is radiated
  collinear to the Born parton $p_\ell$, the observable should behave
  as
\begin{equation}
V({\tilde p}, k)\simeq {d_{\ell}}\left(\frac{k_t}{Q}\right)^{a_\ell}
e^{-{b_{\ell}}\eta}
{g_{\ell}(\phi)}\>, 
\end{equation}
where, for each hard leg $\ell$, $a_\ell, b_\ell, d_\ell$ are some
numbers and $g_\ell(\phi)$ is a regular function parameterising the
azimuthal dependence.

\item The observable should be {\em continuously global}, meaning that
  it should be {\em sensitive to emissions everywhere} and the {\em
    transverse momentum dependence should be uniform ($a_1 = \dots
    =a_n=a$)}. 
  
\item The observable should be {\em recursive infrared and collinear
    (r-IRC) safe}, meaning that {\em the addition of emissions which
    are much softer or more collinear should not drastically change
    the value of the observable}. This new concept was introduced and
  illustrated in detail~\cite{Caesar}.
\end{itemize}
Given the general master formula, together with it's applicability
conditions, \caesar works as an expert system, which in a first step
establishes whether the observable is within its scope. It's next task
is to determine the inputs needed for the evaluation of the master
formula, basically the coefficients and functions $a_\ell, b_\ell,
d_\ell, g_\ell(\phi)$ together with a NLL correction function $\cF$
which accounts for effects due to multiple
emissions~\cite{Caesar,BSZ01}.  Finally as a last trivial step,
\caesar evaluates the master formula (integrating over different Born
configurations when necessary).

Notice that the first two steps are critical: they require high
precision arithmetic~\cite{MP} to take asymptotic (soft \&
collinear) limits and they follow methods of
{``Experimental Mathematics''}~\cite{ExpMath} to validate
or falsify hypothesis.

Currently implemented processes are 2 \& 3 jets in $e^+e^-$, [1+1] \&
[1+2] jets in $DIS$, Drell-Yan + 1 jet, hadron-hadron dijet events.
Results as obtained with \caesar have been tested against all known
NLL global event shape resummations in $\ee$, DIS and DY.IN hadronic
dijet events no event-shape resummations had ever been carried out, so
there was no standard definitions of event shapes, as in $\ee$ and
DIS.
In the following we will therefore show how to define observables at
hadronic colliders which reconcile the seemingly conflicting
theoretical requirement of globalness and the limited experimental
reach of detectors.
Indeed in hadron-hadron collisions the presence of radiation from the
initial state, together with the limited coverage close to the beam
region (usually expressed in terms of a maximum rapidity
$\eta_{\max}$) makes this a critical issue.
\section{Observables in hadron-hadron dijet events}
\subsection{Directly global observables}
\label{sec:DirGlob}
Measurements for this class of observables include ideally all
particles in the event, i.e. the maximum rapidity $\eta_{\max}$ is
taken as large as experimentally possible.  One then defines variants
of the usual $\ee$ observables , e.\ g.
\begin{itemize}
\item
The transverse thrust
$$
  T_{T} \equiv \max_{\vec n_T} \frac{\sum_i |{\vec
      q}_{\perp i}\cdot {\vec
      n_T}|}{\sum_i q_{\perp i}}\,,
  $$
  where $q_{\perp i}$ are the transverse momenta with respect to
  the Beam axis and $\vec n_T$ defines the transverse thrust axis.
\item 
The thrust minor
$$
T_m=\frac{\sum_i |q_i^{out}|}{\sum_i q_{\perp i}}\,,
$$
here $q_i^{out}$ are the momentum-components out of the event-plane
(i.\ e.\ that containing the beam axis and the $\vec n_T$-axis) .
\end{itemize}
NLL resummations are then valid as long as $\displaystyle \ln(1/v) <
(a+b_{\min}) \eta_{\max} $~\cite{KoutZ0}. Explicit results for the
differential distributions show that most of the data lie in this
region~\cite{BSZhh}.

\subsection{Observables with recoil term}
The drawback of observables presented in Sec.~\ref{sec:DirGlob} is
that they require measurements quite far in the forward region. While
this is possible at the Tevatron (where $\eta_{\max}\sim 3.5$), only
calorimeter information is available in forward regions, whose
resolution is quite low.
A way to circumvent this is to define measurements which explicitly
select a restricted, central region and to modify the observable's
definition so as to make it indirectly sensitive to emissions in the
unobserved phase space region. This sensitivity is typically achieved
exploiting recoil effects.
The idea is essentially that momentum conservation ensures that the
{\em vectorial} sum of transverse momenta in the central region,
exactly balances the vectorial sum of transverse momenta in the
unobserved forward regions.
The definition of this type of observables proceeds then as follows.
One first selects a central region $\cC$ (e.\ g.\ taking all particles
with $\eta < \eta_0 \sim 1$), one then defines a central, non-global
observable, e.\ g.\ a central thrust minor
\begin{equation}
T_{m,\cC} \equiv \frac{1}{Q_{\perp,\cC}} \sum_{i \in \cC} |q_{i}^\mathrm{out}|\>, 
\qquad Q_{\perp,\cC} = \sum_{i \in \cC} |\vec q_{\perp i}|\>,
\end{equation}
and adds to it a recoil term, e.\ g. 
\begin{equation}
\cR_{\perp,\cC} \equiv \frac{1}{Q_{\perp,\cC}} \left|\sum_{i \in \cC} {\vec
        q}_{\perp i}\right|\>. 
\end{equation}
The recoil enhanced thrust minor is then given by 
\begin{equation}
  T_{m,\cR}\equiv\, T_{m,\cC} + \cR_{\perp,\cC}\>. 
\end{equation}
These observables are often named indirectly global observables,
since they are indirectly sensitive of emissions in the unobserved
region through recoil effects.
The drawback of these observables is that in the region of extremely
small $V$ the NLL function $\cF$, parameterising multiple emission
effects~\cite{BSZ01}, has a divergence.
The presence of this divergence is well understood, and is not
specific of indirectly global observables.  A divergence occurs
whenever, due to cancellation between contributions from different
emissions, an observable can be very small in the presence of
radiation.
The origin of the divergence can be understood with simple arguments:
the master resummation formula assumes that the mechanism responsible
for keeping the value of the observable small is a LL Sudakov effect.
However, in the phase space region of very small $V$, it is more likely
that $V$ is small because of bi-dimensional cancellations in
$\cR_{\perp,\cC}$. The NLL function $\cF$ cannot compensate a wrong LL
behaviour and manifests this with a divergence.
Despite the presence of the divergence, explicit results show that
most of the distribution is in a region which is under control of NLL
resummations~\cite{BSZhh}.
What makes these observables challenging experimentally, is the
accuracy with which the recoil term can be measured, indeed
$\cR_{\perp,\cC}$ is subject to big cancellations between almost equal
and opposite large transverse momenta.
\subsection{Observables with exponentially suppressed forward terms}
\label{sec:ExpSup}
A variant of directly global observables is to define a class of
observables with exponentially suppressed forward terms.

One introduces the mean transverse-energy weighted rapidity
  $\eta_\cC$ and total transverse momentum of a given  central region $\cC$
\begin{equation}
  \label{eq:etabar}
  \qquad \qquad \eta_\cC = \frac{1}{
    Q_{\perp,\cC}} \sum_{i\in\cC} \eta_i\, q_{\perp i}\>,\qquad\quad
    Q_{\perp,\cC} = \sum_{i\in \cC} q_{\perp i}\>, 
\end{equation}
and with particle in the forward region one defines an
exponentially suppressed forward term
\begin{equation}
  \cE_\barcC = \frac{1 }{Q_{\perp,\cC}}
  \sum_{i \notin \cC} q_{\perp i} \,e^{-|\eta_i - \eta_\cC|}\>, 
\end{equation}
then e.\ g.\ the thrust minor with exponential forward suppression is 
\begin{equation}
T_{m,\cE} = T_{m,\cC} + \cE_{\barcC}\quad\>.
\end{equation}
These observables are intended to have some of the better features of
both purely global and indirectly global observables.  Theoretical
predictions are not affected by divergences and from an experimental
side these observables are optimal in that there is no need for a fine
resolution in rapidity and azimuth in the forward region, and
calorimeter information should be enough to determine the forward term
$\cE_{\barcC}$.
\section{Final considerations}
Despite the fact that observables with exponentially suppressed
forward terms seem to be more suitable in many respects, we point out
that
observables in different classes have complementary sensitivities to
perturbative and non-perturbative (NP) radiation, so that the {\em
  simultaneous study} of a number of observables within each class is
a powerful tool to investigate properties of QCD radiation, e.\ g.\ 
for
\begin{itemize}
\item {\it studies of underlying event}, since the forward sensitivity
  (to beam-fragmentation) can be specifically tuned,
  suppressing it for purely PT studies, or deliberately
  enhancing it when studies NP effects.
    This allows one to test quantitative, as well as qualitative
  features of existing models, e.\ g.\ the sensitivity of the
  underlying event to the partonic channel (i.e. $qq \to qq$, $qg\to
  qg$ or $gg \to gg$).
\item {\it studies of hadronization corrections in multi-jet events},
  in particular for tests of power-corrections beyond the
  ``Feynman Tube model''~\cite{tube-model,ShapeFunctions}. 
\item {\it studies of the non-trivial quantum evolution of colour},
  indeed the novel perturbative QCD colour evolution structure that
  arises in events with 4-jet topology~\cite{BottsSterman} has never
  been investigated experimentally before.
\end{itemize}
Given the much more challenging environment in hadronic collisions,
compared to $\ee$ and DIS, we believe that only an automated approach
can make such a programme feasible.
A number of other possible measurements are given in~\cite{BSZhh},
while results for the automated analysis and resummation of a large
number of observables are available from http://qcd-caesar.org~.
\section{Acknowledgements}
I wish to thank the organisers of ISMD2004 for the invitation. This
work was carried out together with Andrea Banfi and Gavin Salam.

\end{document}